# Planar photonic crystals for express analysis of liquids

Sergey Polevoy [1,*], Arthur Vakula [1,2], Sergey Nedukh [1,3], Sergey Tarapov [1,2,3]

[1] O.Ya. Usikov Institute for Radiophysics and Electronics of NAS of Ukraine, Kharkiv, Ukraine,
[2] Kharkiv National University of Radio Electronics, Kharkiv, Ukraine
[3] Karazin Kharkiv National University, Kharkiv, Ukraine
[*] polevoy@ire.kharkov.ua

**Abstract**

The metamaterial based on two planar photonic crystals was used on experimental express analysis of liquids. The metamaterial operates at a frequency of about 9.5 GHz, and has a small size. It was shown experimentally that when the liquid in container was placed in the resonance region at the interface of the two photonic crystals the parameters of resonance peak of transmission coefficient were changed. It was shown experimentally that different liquids have a different character of depending of the inverse $Q$-factor and the resonance frequency of the resonance peak on the distance from the interface of the photonic crystals to the container with liquid, and this dependence leads to one-to-one identification of such liquid. Thus, the ability of express analysis of liquids in a container by using a metamaterial based on two planar photonic crystals is demonstrated experimentally.

## 1   Introduction

At present, the express analysis of the liquid placed the container is an actual task. Such analysis is required for the fast non-contact identification of liquids, which can be used, for example, in airports in order to filter out liquids in bottles that are forbidden to transportation. Some devices aimed on this purpose are known today. They are based on small perturbation theory [1], which was successfully implemented for instance, in [2], where the cylindrically shaped dielectric resonators are described. They operate at frequencies of about 2 GHz, and have a relatively large size. Therefore, the miniaturization of such devices up to the pocket size is of quite importance. In the present work the metamaterial based on two planar photonic crystals [3] has been applied for the experimental express-analysis of liquids at the resonance frequency about 9.5 GHz.

## 2   Experiment and data analysis

The physical base of the technique is the following. As it was shown earlier [3, 4], metamaterial based on two planar photonic crystals with some certain parameters possess a set of passbands and stopbands in the spectrum. The narrow resonance transmission peak, identified as electrodynamic analog of "Tamm" state, appears in one of stopbands. For the formation of the peak in the transmittance spectrum of the metamaterial the effective resistance for the two photonic crystals at a certain frequency must be equal, and this frequency must be within the stopbands of both photonic crystals. At the frequency of the peak the electromagnetic energy concentration occurs in the vicinity of the interface of two photonic crystals [3]. The resonance field is partially illuminated to the ambient space.

The experimental setup for the study of planar metamaterials is shown in Fig. 1 (b). It consists of the metamaterial under study in microstrip design, connected by a coaxial-microstrip adapters and coaxial cables with a vector network analyzer Agilent N5230A for the measuring of the transmission coefficient of electromagnetic waves in the frequency range 9-10 GHz. Two-coordinate scanning device was used for precise positioning of the measured liquid in the container relative to the interface of two photonic crystals [5]. It allows to set the coordinates of the container by a computer with precision less than 0.1 mm.

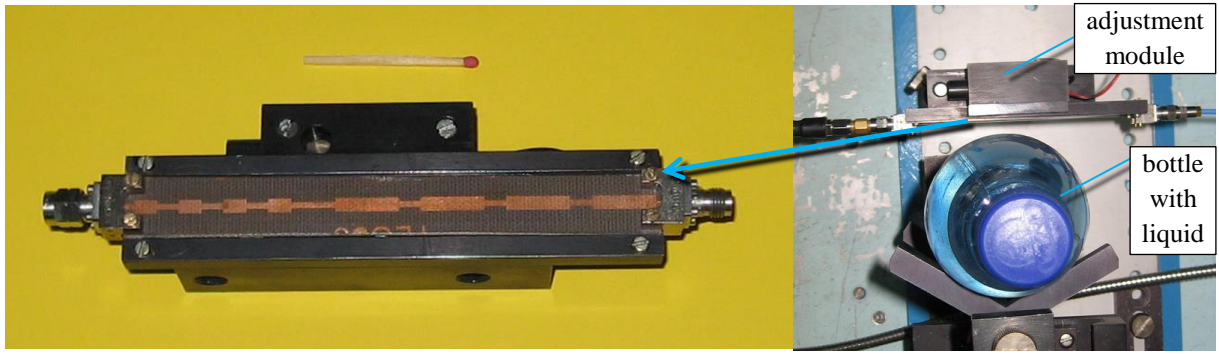

(a)

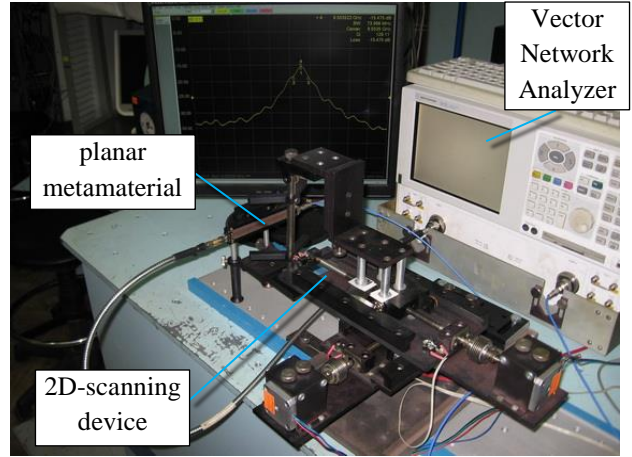

(b)

Figure 1: (a) Photo of metamaterial based on two planar photonic crystals for experimental express analysis of liquids; (b) photo of the experimental setup

The additional adjustment module is a system including the LED laser, optical divider element and a photodetector which are arranged in the axis along the side of the metamaterial. An important condition is to place the dividing element in the area of $E$-component of electromagnetic wave maximum. The technique of adjustment is the following. At the moment when the circular cylindrical container is centered, the intensity of $E$-component of the electromagnetic wave of laser beam reflected from the surface of the container improves according to the laws of geometrical optics, and the beam moves toward the divider. By passing through the dividing element, the laser beam hits the photodetector, which detects a maximum signal from the beam. In the case where container is shifted, the laser beam is declined and does not reach the photodetector. Thus the signal from the beam falls to zero. This design allows to localize fully measured object (possessing the circular cylindrical shape) in space.

The resonance peak frequency and its $Q$-factor are determined by measured frequency dependence of the transmission coefficient of the electromagnetic wave as follows. First, the least-squares method of approximation of the transmission coefficient spectrum by Lorenz curve shape is carried out, which has the form

$$T = c + \frac{a}{1 + b \cdot (f - f_{res})^2} \qquad (1)$$

where $a$, $b$, $c$, $f_{res}$ are adjustable parameters. The parameters of the resonance peak are defined by obtained approximation dependence: the resonant frequency $f_{res}$, $Q$-factor $Q = f_{res}\sqrt{b \cdot (1+\sqrt{2})}/2$ for the width of the curve at the level $1/\sqrt{2}$ from the maximum value.

During the experiment the parameters of peak, such as resonance frequency and the inverse quality factor are determined. It was shown experimentally that when the liquid in container was placed in the resonance region at the interface of two planar photonic crystals the parameters of resonance peak (the "Tamm" peak) are varying. At fixed distance from the interface of the photonic crystals to the container (the bottle) with liquid, we have only one (a single) point in the graph with coordinates of the peak resonance frequency $f_{res}$ and inverse $Q$-factor [2]. When this distance $\delta h_{var}$ is changed, we obtain a set of these peak parameters. It is suitable to present them as some certain curve on a graph (Fig. 2).

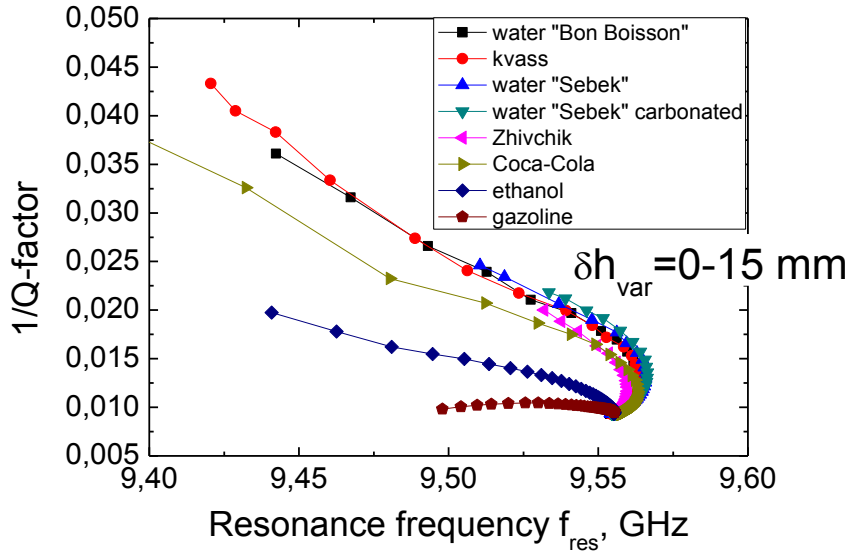

Figure 2: The inverse Q-factor and the resonance frequency for the various distances from the interface of the photonic crystals to the container and for the different liquids in bottle

It was shown experimentally the different character of dependence of the inverse *Q*-factor and the resonance frequency $f_{res}$ on the distance $\delta h_{var}$ for the different liquids in bottle (Fig. 2). Note that the large losses in the liquid lead to peak *Q*-factor decreasing, and the above-described curve is shifted up. The liquids based on water solutions with a large absorption have the highest reverse quality factor. Ethyl alcohol has the intermediate values, and gasoline is almost no effect on the reverse quality factor of the peak due to small losses. Thus, rapid express analysis of liquid may consist in the selection of such previously measured curve for the known liquid which is most closely approaches the measuring of the unknown liquid, which leads to its identification. To improve the reliability of liquid identification it is possible to carry out measurements at multiple frequencies, that is possible for such type of metamaterials [4].

The investigation of the influence of the ethyl alcohol concentration in water to the shape of dependence of the inverse *Q*-factor and resonance frequency on the distance $\delta h_{var}$ was carried out (Fig. 3). It can be seen that dependences on the figure are clearly distinguishable, which makes it possible to estimate the alcohol concentration in solutions on the express analysis of liquids.

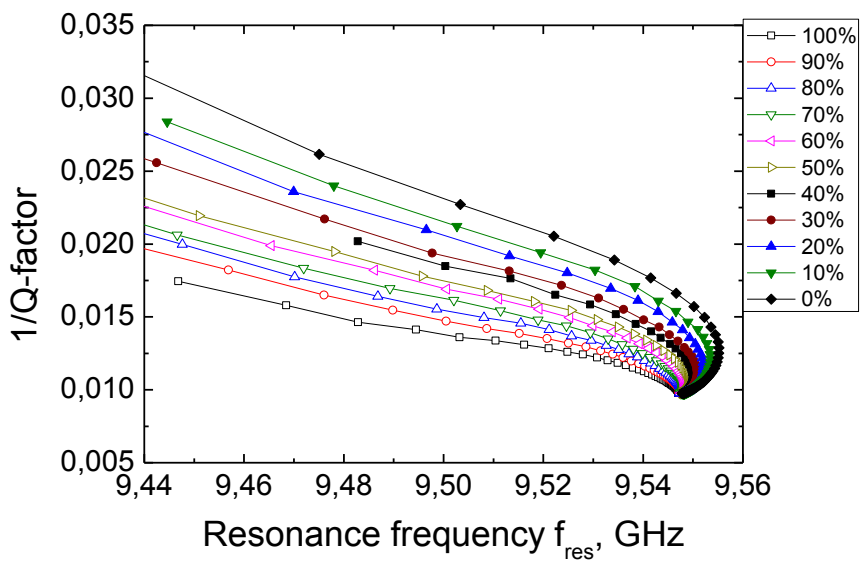

Figure 3: The inverse Q-factor and the resonance frequency for the various distances from the interface of the photonic crystals to the container and for the different concentration of ethanol in water

## 3    Conclusions

Thus, the technique for express-analysis and control of liquids is developed. The technique is presented at frequency of about 9.5 GHz; thus the device which has been designed for this technique has quite small size. It was shown experimentally the different character of dependence of the inverse $Q$-factor and the resonance frequency $f_{res}$ on the distance $\delta h_{var}$ from the interface of the photonic crystals to the different liquids in bottle. The rapid express analysis of liquid may consist in the selection of such previously measured dependence for the known liquid which is most closely approaches the measuring of the unknown liquid, which leads to its identification. The advantages of technique based on the metamaterial formed by two planar photonic crystals is demonstrated experimentally.